\documentclass[twocolumn,prd,aps,amsmath,amssymb,epsfig]{revtex4}

\usepackage{graphicx}
\usepackage{dcolumn}
\usepackage{bm}

\begin{document}
\title{${\cal N}=1$ Super Yang Mills renormalization schemes for
Fractional Branes}
\author{Raffaele {\sc Marotta}$^a$ \quad and \quad Francesco {\sc Sannino}$^b$}
\address{\mbox{$^a${Dipartimento~di~Scienze~Fisiche,~Universit\`a~di~Napoli~and~INFN,~Sezione~di~Napoli}},\mbox{Via~Cintia
Complesso~Universitario~M.~Sant'~Angelo~I-80126~Napoli,~Italy}\\~\\\mbox{$^b${\rm
NORDITA},~Blegdamsvej~17,~DK-2100~Copenhagen~\O,~Denmark}}
\date{July 2002}

\begin{abstract}
We investigate ${\cal N}=1$ super Yang-Mills theory using
fractional branes. We first define the $\beta$-function with
respect to a supergravity coordinate. To provide the relation
between the supergravity parameter and the renormalization group
scale we use the UV known gauge theory $\beta$-function as a type
of boundary condition. We show that there are no privileged
renormalization schemes connected to a given supergravity solution
while we investigate  in some detail two schemes. The Wilsonian
one where just one loop is manifest and the one containing
multi-loops. A new functional relation between the gaugino
condensate and the supergravity coordinates is finally determined.
\end{abstract}

\maketitle

\section{Introduction}
\label{introduction}

Recently using the gauge/gravity correspondence correspondence
it was shown that many gauge theories can be investigated via
their dual supergravity backgrounds. In the case of D3-branes in
flat space this correspondence became even more stringent since an
exact duality between ${\cal N}=4$ super Yang-Mills in four
dimensions and type IIB string theory on $AdS_5\times
S_5$\cite{Maldacena} is conjectured to exist.

Clearly one would like to extend such a duality to less
supersymmetric and non conformal gauge-theories. {}Fractional
branes are a tool to construct non-conformal gauge theories with
reduced supersimmetry\cite{BDM}-\cite{B}. More specifically by
using fractional branes at orbifold singularities it has been
possible to reproduce some gauge theory 1-loop perturbative
results such as the $\beta$-function and the chiral  anomaly.
These results are not interpreted with a duality \'a la Maldacena,
but are consistently interpreted as a open/closed string
duality\cite{Bertolini:2001qa}. The classical supergravity
solution, indeed, can be seen as an expansion for small 't Hooft
coupling and as such it is not surprising that one is able to
deduce the perturbative properties of the gauge theory. When this
approach has been extended to describe fractional branes
supporting in their world-volume ${\cal N}=1$ super Yang-Mills
theories, one seems able to deduce just the 1-loop Wilsonian
$\beta$-function\cite{BDFM}.

{}On the other hand for the ${\cal N}=1$ Yang-Mills theory it has
been shown that there exists a precise relation between the 1-loop
Wilsonian $\beta$-function and the  multi loop NSVZ one computed
in a non holomorphic (in the coupling constant) scheme
\cite{Shifman:1999kf}. Having at hand such a relation it is
reasonable to expect that the two schemes contain the same
physical information.

Since it is believed valid the supergravity-gauge theory
correspondence one should then be able to reproduce also the
running of the coupling constant in this non-holomorphic
renormalization scheme within the framework of the singular
supergravity solutions. One of the main goal of this paper is to
propose a solution to this problem.

In  particular we will first observe that there is a good deal of
arbitrariness when trying to identify the supergravity parameters
(in this case the coordinates transverse to the brane) to the
4-dimensional renormalization scale of the theory. This
arbitrariness is essentially due to the possibility of choosing in
non-conformal backgrounds, different functional relations between
distance in supergravity and energy scale in gauge theory
\cite{PP}.
 We will then
show how this arbitrariness can be used to go from one
renormalization scheme to another.

We will also provide a new functional relation between a specific
combination of the supergravity coordinates and the gaugino
condensate. We will use again the UV boundary conditions to
completely determine this function at large scales.

Since the specific supergravity solution we are considering is
singular we will not attempt to understand the IR of the
corresponding gauge theory. This has been explored in
\cite{{DLM},I,PF} using non singular supergravity solutions
\cite{MN,{Klebanov:2000hb}}. In particular in Ref.~\cite{PF} it
has been shown that the UV gauge theory boundary conditions are
relevant to understand the non perturbative IR physics of super
Yang-Mills.

In section {\ref{SYM}} we will briefly review some of the relevant
aspects of ${\cal N}=1$ Super Yang-Mills theory. In section
{\ref{fractional}} we will introduce the fractional brane set up
while in section \ref{UVBC} we define and impose the UV boundary
conditions. The new relation between the gaugino condensate and
the supergravity coordinates is introduced and studied in section
\ref{gluino}. The conclusions are presented in \ref{Finale}.

\section{Super Yang Mills perturbative $\beta$-functions}
\label{SYM} In this section we summarize same general feature of
pure $SU(N)$ super Yang-Mills theory. At the classical level the
theory possesses a $U(1)_R$ symmetry which does not commute with
the supersymmetric algebra. The ABJ anomaly breaks the $U(1)_R$
symmetry to $Z_{2N}$. Non perturbative effects trigger the gaugino
condensation leading to further breaking of $Z_{2N}$ symmetry to a
left over $Z_2$ symmetry and we have $N$ equivalent vacua. The
gaugino condensate is a relevant ingredient to constrain the
theory. In literature it has been computed in different ways and
we recall here its expression in two much studied schemes
\cite{Shifman:1999kf,HKLM}:
\begin{itemize}
\item{\underline{The Holomorphic Scheme/Wilsonian}
\begin{eqnarray}
\langle \lambda^2  \rangle &=&Const.\, \mu^3 e^{i\frac{2\pi}{N}
{\tau_{\rm YM}}} \nonumber \\ &=& Const.\, \mu^3 \, e^{
-\frac{8\pi^2}{N\, g^2_{\rm YM}}} \,\,\, e^{ i\frac{\theta}{N}}
\label{ino1}
\end{eqnarray}
where the gaugino condensate is holomorphic in $\tau_{\rm YM}=
\frac{\theta}{2\pi} + i \frac{4\pi}{g^2_{YM}}$.}
\item{\underline{The Non Holomorphic Scheme/Pauli Villars}
\begin{eqnarray}
\langle \lambda^2  \rangle &=& Const.\, \mu^3
Im\left[\frac{\tau_{\rm YM}}{4\pi}\right]e^{ i\frac{2\pi}{N}
{\tau_{\rm YM}}} \nonumber \\ &=& Const.\, \mu^3 \frac{1}{g^2_{\rm
YM}} e^{ -\frac{8\pi^2}{N\, g^2_{\rm YM}}} e^{i \frac{\theta}{N}}
\label{gnh}
\end{eqnarray}
Here the condensate is non-holomorphic in $\tau_{\rm YM}$.}
\end{itemize}
The gaugino condensate is a universal physical constant
independent on the scheme. This fact not only allows one to
compute the {\it perturbative} $\beta$-functions in the two
schemes since the respective coupling constant must depend on the
scale in such a way to compensate the dependence on $\mu$ but it
also enables us to establish a relation between the coupling
constants in the two schemes.

The holomorphic scheme is very constrained leading to a pure
one-loop type of running. This is a welcome feature from the point
of view of the supercurrent chiral multiplet
\cite{Shifman:1999kf}. The second scheme is non holomorphic in the
coupling constant. The theta dependence is constrained by the
$U(1)_R$ anomalous symmetry. The two coupling constants are
related in the following way:
\begin{eqnarray}
{\tau_{\rm YM}}^{H}=\tau_{\rm YM}-i\frac{N}{2\pi}\ln \left[{\rm
Im}\left( \frac{\tau_{\rm YM}}{4\pi}\right)\right] \ ,
\label{coupling-relation}
\end{eqnarray}
where we added a superscript $H$ to distinguish the holomorphic
coupling from the non holomorphic one. Note that the transition
from one to the other scheme does not alter the theta dependence
of the theory. This request manifestly breaks the holomorphicity
in the coupling constant for the non Wilsonian scheme.  The
independence on the scale of the gaugino condensate leads to the
following $\beta$-functions \cite{Shifman:1999kf}:
\begin{eqnarray}
\beta \left(g_{\rm YM}\right)&=& -\frac{3N}{16\pi^2}{g^3_{\rm YM}}, \qquad {\rm Holomorphic}  \\
\beta\left(g_{\rm YM}\right)&=& -\frac{3N}{16\pi^2}\,{g^3_{\rm
YM}}\, \left[{1-\frac{N g_{\rm YM}^2}{8\pi^2}}\right]^{-1} \ .
\end{eqnarray}
Where $\displaystyle{\beta(g)=\frac{\partial g}{\partial L}}$ and
$\displaystyle{L=\ln (\mu/\Lambda)}$ while $\mu$ is the
renormalization scale and $\Lambda$ a reference scale. Due to the
relation (\ref{coupling-relation}) the one-loop and the
perturbative multi loop $\beta$-function carry the same perturbative
physical information. Although these couplings may capture the all
order perturbative terms the non perturbative contributions (if
any) to the complete $\beta$-functions are still missing.

These two schemes display no universality of the two loop
coefficient of the $\beta$-function. This is due to the non
analytical relation between the coupling constants in the two
schemes (\ref{coupling-relation}). In fact the universality
argument for the two loop $\beta$-function coefficient is valid for
part of the possible choices of renormalization schemes
\cite{Hooft-Book}.

If supergravity-gauge theory correspondence is valid it should be
possible to adopt different renormalization schemes when
connecting the supergravity solutions to the associated gauge
theory. We will address this issue in the following sections.

\section{Fractional Branes: The Set Up}
\label{fractional}
 Let us now consider the  pure SYM realized as
the world-volume theory of  a stack of N D3-fractional branes at
the orbifold ${\rm I\!R}^{1,3}\times
\mathbb{C}^3/(\mathbb{Z}_2\times \mathbb{Z}_2)$. For definiteness
we take the orbifold directions to be $x^4\dots x^9$ (labeled by
$l,m \dots$) introduce three complex coordinates:

\begin{equation}
z_1=x^4+ix^5\,\, z_2=x^6+ix^7\,\,z_3=x^8+ix^9 \label{zorb}
\end{equation}

and consider fractional branes that are completely transverse to
the orbifold, and therefore extended along the directions
$x^0\dots x^3$ (labeled by $\alpha,\beta\dots$ ) .

Within this orbifold one has four kinds of D3-fractional
branes\cite{BDFM}, corresponding to the four irreducible
representations of the orbifold $\mathbb{Z}_2\times \mathbb{Z}_2$,
each of them supporting in its world-volume a ${\cal N}=1$
SYM theory.

In the following we will consider the low energy dynamics of a
bound state made of just one kind  of D3-fractional branes. The
gauge theory supported by a pile of $N$ of such objects is a pure
${\cal N}=1$ super Yang-Mills theory.

The supergravity solution determined by this source has been
computed in the paper of Ref. \cite{BDFM}. In the following we
specify that solution for the case that we are interested in; i.e.
the  case of only one type of fractional brane, obtaining for the
metric and the R-R 4-form $C_4$:
\begin{eqnarray}
ds^2 &=& H^{-1/2} \eta_{\alpha\beta} dx^{\alpha} dx{\beta} +
H^{1/2} \delta_{lm} dx^l dx^m
\label{metric}\\
\tilde{F}_5&=&dH^{-1} \wedge V_4+ *\left( dH^{-1}  \wedge V_4
\right) \label{f5}
\end{eqnarray}
while the classical profile of the twisted  fields is:
\begin{eqnarray}
b_i&=& (2 \pi^2{\alpha'})\; \left(1+\frac{2g_s}{\pi}\,N \, \log
\frac{\rho_i}{\epsilon}\right)\label{bsol1}\\ c_i&=&-
(4\pi\alpha')\,g_s\,N\,\theta_i \label{csol1} \label{metric1}
\end{eqnarray}
with $\rho_i = \sqrt{z_i \bar{z}_i}$, $\theta_i={\rm
tan}^{-1}(x^{2i+3}/x^{2i+2})$  $(i=1,2,3)$.
The  three pair of scalars $b_i$ and $c_i$ ($i=1,2,3$) correspond
to the components of the 2-forms $B_{(2)}$ and $C_{(2)}$, along
the anti-self dual 2-forms $\omega_{(2)}^i$ dual to the cycle
${\cal C}_i$ that caracterize this orbifold.

The gauge coupling and the theta angle of the dual gauge theory,
are given in terms the twisted fields fluxes by the following
relations \cite{BDFM}:
\begin{eqnarray}
\frac{1}{g^2_{YM}} &=& \frac{1}{16\pi
g_s}+\frac{N}{8\pi^2}\,\sum_{i}
\ln\frac{\rho_i}{\epsilon}\nonumber\\
&&\label{par0}\\
\theta_{\rm YM} &=&-N\sum_i\,\theta_i \ . \label{par1}
\end{eqnarray}

Combining the previous equations we have
that the complex coupling $\tau_{\rm YM}$ is:
\begin{eqnarray}
&&\tau_{\rm YM} = {\rm i}\left[
\frac{1}{4g_s}+\frac{N}{2\pi}\sum_i \ln\frac{z_i}{\epsilon}\right]
\equiv {\rm i}\frac{N}{2\pi}\sum_i
\ln\frac{z_i}{\langle z \rangle} \ ,\nonumber\\
&&\label{tau11}
\end{eqnarray}
where $\langle z \rangle $ is nothing but $\langle \rho \rangle$
when restricting ourselves to the real part. $\langle z \rangle $
is real since we have chosen in eq.(\ref{metric1}) a null
background value for the $c_i$ fields. However while still
satisfying the equation of motion we can make a different constant
choice for such background which promotes $\langle z \rangle$ to a
complex variable.

 It is always possible and natural to introduce
the following dimensionless variable
\begin{eqnarray}
\rho^3 \equiv \prod_{i=1}^3\frac{\rho_i}{|\langle z  \rangle |} \
,\label{rho}
\end{eqnarray}
with respect to which the functional relation between the coupling
constant reads:
\begin{eqnarray}
\frac{1}{g_{\rm YM}^2}= \frac{3N}{8\pi^2}{\ln \rho}\equiv F(\rho)
\ . \label{radialef}
\end{eqnarray}
Being the supergravity solution singular we cannot explore the IR
of the theory which would correspond to small values of $\rho$.
This is possible, as already discussed, in the case of the wrapped
brane scenario. The hope is that, in a not too far future, by
resolving the singularity one can check the super Yang-Mills IR
physics using fractional branes.

It is worth mentioning that it has been very useful in the wrapped
brane scenario to relate a known function of the supergravity
coordinates with the gaugino condensate \cite{ABCPZ}.  We shall
suggest, in the last paragraph, how such a relation may emerge in
the fractional brane scenario as well.

\section{UV Boundary Conditions:  From One Loop to Multi Loops}
\label{UVBC} Supergravity provided the following functional
relation:
\begin{eqnarray}
\frac{1}{g^2_{\rm YM}}={F}(\rho), \qquad \rho\rightarrow \infty \
.
\end{eqnarray}
In order to compute the standard gauge theory $\beta$-function we
must differentiate the coupling constant with respect to $L=\ln
\left(\mu/\Lambda\right)$ with $\mu$ the renormalization scale and
$\Lambda$ a renormalization invariant scale. Supergravity  is not
able to predict such a relation between $\rho$ and $\mu$. Without
loss of generality we can nevertheless define the following
unambiguous supergravity $\beta$-function for fractional branes:
\begin{eqnarray}
\beta_{\rho}\left(g_{\rm YM}\right)=\frac{\partial g_{\rm
YM}}{\partial \rho} =-\frac{3N}{16\pi^2}\, g^3_{\rm YM}\,
\frac{1}{ \rho}, \qquad \rho\rightarrow \infty ,
\end{eqnarray}
which clearly captures asymptotic freedom. Once the relation
between $\rho$ and $\mu$ is known the standard $\beta$-function is
simply:
\begin{eqnarray}
\beta\left(g_{\rm YM}\right) = \beta_{\rho}\left(g_{\rm YM}\right)
\frac{\partial\rho}{\partial L}.
\end{eqnarray}
For non singular Jacobian $\partial \rho/\partial L$ most of the
information carried in $\beta_{\rho}$ is transferred to the gauge
theory $\beta$-function. If this were not the case the
supergravity/gauge connection would be meaningless.

In order to fix the functional dependence between the supergravity
parameter $\rho$ and $L$ we use the known form of the
$\beta$-function in UV. This UV type of boundary conditions are
similar to the ones introduced in the context of the wrapped brane
scenario \cite{PF}.

\subsection{Wilsonian UV Boundary Condition}
To reproduce the $\beta$-function in this scheme the UV boundary
condition reads:
\begin{eqnarray}
\frac{\partial \ln \rho}{\partial L} = 1.
\end{eqnarray}
yielding the following functional relations:
\begin{eqnarray}
\rho=\frac{\mu}{\Lambda}.
\end{eqnarray}
This specific boundary condition reproduces the relation between
$\rho$ and $\mu$ already adopted in literature.

\subsection{Pauli Villars UV Boundary Condition}
We have already stressed that the physical information encoded in
the $\beta$-function expressed in this scheme is the same than the
one carried in the Wilsonian scheme so there is no reason for the
supergravity solution not to be linked to this scheme as well.
Here the UV boundary condition is:
\begin{eqnarray}
\frac{\partial y(\rho)}{\partial L} &=& \left[1 - N \frac{g^2_{\rm
YM}}{8 \pi^2}\right]^{-1} \nonumber \\
&=& \left[1 - \frac{1}{3y}\right]^{-1} , \label{2boundary}
\end{eqnarray}
with $y(\rho)=\ln \rho$ and to write the last identity we have
used eq.~(\ref{radialef}). After integrating  we have:
\begin{eqnarray}
3L=3 \ln {\rho} - \ln \left(\ln{\rho}\right) \ ,
\end{eqnarray}
which can be solved (for large $\mu$) iteratively for $\rho$
yielding:
\begin{eqnarray}
\rho \approx \frac{\mu}{\Lambda} \sqrt[3]{\ln\frac{\mu}{\Lambda}},
\qquad \mu \rightarrow \infty \ . \label{rho2}
\end{eqnarray}
Note that in this scheme the two loop coefficient is not
vanishing. Any other scheme in which the coupling constant
relation can be written as a series expansion with respect to the
new coupling constant will display the universality of the two
loop coefficient of the $\beta$-function \cite{Hooft-Book}. The
Wilsonian coupling relation to the Pauli-Villars scheme is not
analytical and as such does not display the universality of the
loop coefficient of the respective $\beta$-functions. It is also
interesting to observe that our functional relation
eq.~(\ref{rho2}) does not change if we neglect in
eq.~(\ref{2boundary}) terms higher or of the order $g^4_{\rm YM}$.

\section{The hidden gaugino condensate}
\label{gluino}

The gaugino condensate is a key ingredient when studying ${\cal
N}=1$ super Yang-Mills gauge theory. In particular being a
renormalization group invariant quantity is constant for any
(non-zero) value of the coupling constant and in any scheme. In
particular is present at weak coupling.

{}Here we suggest a possible correspondence between supergravity
and the gaugino condensate within the fractional brane scenario.
In order to find such a relation we recall that in supergravity
the combination of twisted fields given by the eq.(\ref{tau11})
is identified with the complex coupling of the gauge theory.

Since under an R-transformation $\tau_{\rm YM}\rightarrow
\tau_{\rm YM}+ \frac{N}{\pi}\alpha$, with $\alpha$ the $U(1)_R$
transformation parameter for consistency
\begin{equation}
\langle z\rangle  \rightarrow  e^{\frac{2}{3}i \alpha}\,
\langle z\rangle \label{mur}
\end{equation}
So by
construction the quantity
\begin{eqnarray}
\prod_{i=1}^3 \langle z\rangle, \label{prodotto}
\end{eqnarray}
has $U(1)_R$ charge two and engineering mass dimension three.
Eq.~(\ref{prodotto}) is also precisely the combination
appearing in eq.~(\ref{tau11}).
Furthermore when {\it choosing} a particular supergravity
scale (i.e. a value for the coordinate $z_i$) we break the R-symmetry
to $Z_2$. It is hence tempting to suggest that there
exists a functional relation between eq.~(\ref{prodotto}) and the
inverse of the gaugino condensate of the form:
\begin{equation}
\prod_{i=1}^3 \left[\frac{\rho_i}{\langle z  \rangle} \right] {\cal
F}\left[{\rho}\right] =\frac{\mu^3}{\langle \lambda^2 \rangle } \
,\label{rel1}
\end{equation}
where the function ${\cal F}$ can be fixed using the UV boundary
conditions used in the previous section or equivalently the
expression for the gaugino condensate presented in
eqs.~(\ref{ino1}), (\ref{gnh}). Equation (\ref{rel1}) shows the
direct link between $\langle z \rangle$ and $\langle
\lambda^2\rangle$ and is consistent with the $U(1)_R$
transformation. Indeed $\langle z \rangle$ can be imagined to be
the vacuum expectation value of $z_i$ with R-charge $2/3$ while
$\langle \lambda^2\rangle$ is the condensate of the gluino which
possesses, in the present conventions, a positive unit of R-charge
while we hold fixed $z_i$ and $\mu$.

Using the gaugino expression in the Wilsonian scheme (see
eq.~(\ref{ino1})) one deduces ${\cal F}={\rm const.}$ while in the
multi loop case (\ref{gnh}) we have ${\cal F}=\displaystyle{{\rm
const.}/{\ln \rho}}$.

It is instructive to determine ${\cal F}$ again, within the
non-holomorphic renormalization scheme, using the universality of
the two loop coefficients of the $\beta$-function. This is
equivalent to the boundary conditions just imposed but it shows
the intimate relation between the gaugino condensate and the beta
function.

By differentiating  both sides of eq. (\ref{rel1}), we get:
\begin{equation}
\frac{\partial}{\partial \log \frac{\mu}{\Lambda}}\ln (\rho)=
\frac{1}{1+ \frac{\rho}{3}\, \partial_{\rho}\ln {\cal F}(\rho)}
\label{rhomu}
\end{equation}
where $\rho$ is defined in eq. (\ref{rho}) and we used the
relation between the gaugino condensate and the dynamical scale
$\Lambda$ of the pure ${\cal N}=1$ theory\cite{HKLM}; i.e. $
\langle \lambda^2 \rangle_{\theta=0} = \Lambda^3$. Furthermore by
using the relations (\ref{rhomu}) and the (\ref{par0}) we can
compute from the classical solution, the $\beta$-function of the
underlying gauge theory expressed in terms of the unknown
function:
\begin{eqnarray}
\beta(g_{\rm YM})&&= - 3 \frac{g_{\rm YM}^3}{16\pi^2} \left[1 +
\frac{\rho}{3}\, \partial_{\rho} \ln {\cal F} (\rho)\right]^{-1}
\nonumber\\
&&= - 3 \frac{g_{\rm YM}^3}{16\pi^2} \left[1 -\frac{\rho}{3}\,
\partial_{\rho} \ln {\cal F}(\rho) \dots \right]
\nonumber\\
\label{beta1}
\end{eqnarray}
The second expression follows because we are considering a
perturbative expansion for the $\beta$-function. Using the
universality of  the first two orders of the $\beta$-function
\cite{Hooft-Book} we deduce (for large $\rho$ i.e. small
coupling constant) the following differential equation:
\begin{equation}
\partial_{\rho}\ln{\cal F}(\rho)=-\partial_{\rho} \ln \ln \rho
\label{relation}
\end{equation}
which fixes $\displaystyle{{\cal F}=\frac{\rm const.}{\ln\rho}}$ in the
UV regime. Substituting this
expression in (\ref{beta1}) yields:
\begin{eqnarray}
\beta(g_{\rm YM})&=& - 3 \frac{g_{\rm YM}^3}{16\pi^2}
\left(1 - \frac{1}{3\, \ln\rho}\right)^{-1}\nonumber\\
&=& - 3 \frac{g_{\rm YM}^3}{16\pi^2} \left( 1 -\frac{N g^2_{\rm
YM}}{8\pi^2} \right)^{-1} \label{beta}
\end{eqnarray}
This is the NSVZ $\beta$-function\cite{Shifman:1999kf}, also found
in Ref.\cite{J}.

\section{Conclusions}
\label{Finale}

We investigated the UV properties of the ${\cal N}=1$ super
Yang-Mills theory using singular supergravity solutions
corresponding to fractional branes. We first defined the
$\beta$-function with respect to the supergravity coordinate
$\rho$ and found a supergravity type of asymptotic freedom. To
provide the relation between the supergravity parameter and the
renormalization group scale we used the UV known gauge theory
$\beta$-function as a kind of boundary condition. The UV boundary
conditions take the form of linear differential equations which
can be easily solved. In particular we have seen that there are no
privileged renormalization schemes connected to a given
supergravity solution. We investigated in some detail two schemes.
The Wilsonian one where, technically, just one loop is present and
the one containing multi-loops. We recall that in the multi-loop
$\beta$-function the first two coefficients are universal
\cite{Hooft-Book} while the others are scheme dependent.

Furthermore, by analyzing how the $U(1)_R$ symmetry is realized in
supergravity, we determined a functional relation between the
gaugino condensate and  a particular combination of the
coordinates transverse to the brane.

The IR physics cannot be uncovered with singular supergravity
solutions. However the hope is that in the future, by resolving
somehow the singularity, one can use fractional branes to get an
independent check of the IR uncovered via wrapped branes
\cite{DLM}$-$\cite{PF}. Besides, at the moment, within the
fractional brane scenario it is possible to investigate ${\cal
N}=1$ supersymmetric gauge theories with matter fields.

\acknowledgments We thank L. Cappiello, F. Cuomo, P. Di Vecchia,
E. Imeroni, A. Lerda, A. Liccardo, R. Musto, R. Pettorino and F.
Pezzella for discussions. We also thank F. Nicodemi and W. M\"uck
for a careful reading of the manuscript while P. Olesen for
enlightening discussions, encouragements and a careful reading of
the manuscript. R.M. would like to thank Nordita for the kind
hospitality. The work of R.M. is partially supported by EC program
HPRN-EC-2000-00131 and MIUR while F.S. is supported by the
Marie--Curie fellowship under contract MCFI-2001-00181.

\end{document}